\documentclass[fleqn,usenatbib]{mnras}

\usepackage{newtxtext,newtxmath}

\usepackage[T1]{fontenc}

\DeclareRobustCommand{\VAN}[3]{#2}
\let\VANthebibliography\thebibliography
\def\thebibliography{\DeclareRobustCommand{\VAN}[3]{##3}\VANthebibliography}

\usepackage{graphicx}	% Including figure files
\usepackage{amsmath}	% Advanced maths commands
\usepackage{amssymb}	% Extra maths symbols
\usepackage{longtable}
\usepackage{soul}
\usepackage{tikz}
\usepackage{textcomp}
\usepackage{multirow}
\usepackage{caption}
\usepackage{subcaption}
\usepackage{caption}
\usepackage{array}
\usepackage{hyperref}
\usepackage{natbib}
\usepackage{diagbox}
\usepackage{booktabs}
\title[Testing unification models]{Unified models revisited I: modelling the effect of source geometry on radio galaxy/quasar unification}

\author[S. Pinjarkar et al.]{
	Siddhant Pinjarkar$^{1}$\thanks{E-mail: s.pinjarkar@herts.ac.uk},
	Martin J.~Hardcastle$^{1}$, 
    Jonathon C.S. Pierce$^{1}$, and
    Frits Sweijen$^{2}$
	\\
	$^{1}$Centre for Astrophysics Research, Department of Physics, Astronomy and Mathematics, University of Hertfordshire, College Lane, Hatfield AL10 9AB, UK\\
    $^{2}$Centre for Extragalactic Astronomy, Department of Physics, Durham University, Durham DH1 3LE, UK\\}

\date{Accepted XXX. Received YYY; in original form ZZZ}

\pubyear{2024}

\begin{document}
\label{firstpage}
\pagerange{\pageref{firstpage}--\pageref{lastpage}}
\maketitle

\begin{abstract}
The orientation-based unification model proposes that radio-loud quasars and radio galaxies are the same objects observed at different angles. A key prediction of this model is that the quasars are seen at smaller angles to the line of sight and so should be more affected by projection, and hence apparently physically smaller, than corresponding radio galaxies, but this has not always been found in earlier studies. We argue that the interpretation of observations requires a less simplistic model for the effects of projection, which takes into account radio sources' finite width and their intrinsic axial ratio distribution. Using this cylindrical configuration as a basis for the simulation of radio galaxies and quasars, we simulate the distribution of the linear size ratio of quasars to radio galaxies for different sample sizes and critical angles. Our simulations that predict the ratio of observed lengths in the presence of a distribution of intrinsic physical sizes and axial ratios that we derive from observation. We conclude that to test the unified scheme, samples should be completely optically identified, sizes should be measurable for all targets, and the sample size should be greater than $\sim 800$ sources. Such large samples with uniform optical identification and accurate size measurements have not been available in previous work, but should become available from wide-area sky surveys in the near future. 
\end{abstract}


%%%%%%%%%%%%%%%%%%%%%%%%%%%%%%%%%%%%%%%%%%%%%%%%%%

%%%%%%%%%%%%%%%%% BODY OF PAPER %%%%%%%%%%%%%%%%%%

\section{Introduction}
\label{Intro}
\subsection{Quasars and radio galaxies}

The ultimate source of the luminosity of Active Galactic Nuclei (AGN) is accretion onto the central supermassive black hole. For radiatively efficient (RE) objects \citep{Hardcastleetal2018} this is thought to take place through a geometrically thin accretion disk\footnote{There are many reasons to believe that the standard picture of these disks is lacking important physics \citep{Antonuccietal2023}; here we use the term `disk' simply as a shorthand for the geometrically thin structure that seems likely to exist without subscribing to any particular model for the origin of the continuum emission.}. Dissipation in the disk leads to continuum radiation in the optical through UV and, via Compton scattering, the X-ray. The hard radiation field ionizes material both close to and far away from the central black hole, giving rise to the broad-line region (BLR) and narrow-line region (NLR), respectively.

This picture is more or less independent of the orientation of the disk with respect to the line of sight, but it has been known for some time that there exist `type II' AGN which have a narrow-line region of a luminosity comparable to those seen in normal, or `type I', AGN, but do not have the strong optical/UV continuum emission or broad-line emission of the type Is. The generally accepted explanation \citep{Antonuccietal1993} is that an anisotropic obscuring structure, the `torus', surrounds the accretion disk, obscuring a direct view of the continuum emission and BLR from certain lines of sight \citep{Kormendyetal1995,Magorrianetal1998,Ferrareseetal2000} while the NLR largely lies on larger scales than the torus and so is seen from all viewing angles. The clinching argument for this picture originally was the detection of the BLR in polarized emission in type II objects, showing that it is present but not directly observable \citep{Antonuccietal1984,Antonuccietal1985}. Since then the strong obscuration seen in the X-ray spectrum of type II objects and the detection of mid-IR emission consistent with obscuration via the torus have provided additional support to this `orientation-based unification' picture \citep{Whysongetal2004, Ogleetal2006, Hardcastleetal2009, Netzeretal2015}.

AGN with strong jet-driven radio emission (hereafter radio-loud AGN, or RLAGN) add a complication but also allow additional tests of orientation-based unified models. While some RLAGN behave as more or less standard type I or type II AGN with the addition of strong radio emission (the so-called narrow-line radio galaxies, NLRG, being type II and radio-loud quasars being type I) there exists a numerically dominant population of what are commonly known as low-excitation radio galaxies (LERG), which have been known for many years \citep{Hineetal1979}. These objects do not show any of the apparatus of the type I/II standard AGN, including, crucially, no evidence for an obscuring torus that might hide it \citep{Whysongetal2004,Hardcastleetal2009}. LERGs are thought to accrete through a geometrically thick and optically thin, radiatively inefficient accretion disk \citep[e.g.][]{Narayanetal1995}, producing very little radiation while still having potentially highly energetic radio jets, and it now seems clear that these `radiatively inefficient' AGN are the result of accretion below some critical Eddington-scaled accretion rate threshold \citep{Bestetal2012,Mingoetal2014}. LERGs must therefore be excluded from any consideration of a radio-loud orientation-based unified model \citep{Hardcastleetal1999}.

If LERGs are excluded, though, RLAGN have a number of advantages in tests of unified models. Firstly, low-frequency radio emission is (almost) isotropic, and so allows selection of samples that are unbiased with respect to orientation. In addition, AGN jets are relativistically beamed \citep{Laingetal1988,Garringtonetal1988} and appear to emerge more or less perpendicular to the torus, so that quasars show stronger, more one-sided jets, apparently faster bulk motions on parsec scales, and more prominent cores than NLRG. These key facts led \cite{Barthel1989} to propose the original radio-loud orientation-based unified model, with a critical angle of $\sim 45^\circ$ to the line of sight separating quasars and NLRG. \cite{Barthel1989} showed that the numbers of 3CRR radio galaxies and quasars with $0.5 < z < 1.0$ were consistent with a critical angle of $\sim 45^\circ$, and argued that the fact that the quasars in the sample were apparently smaller in size than the radio galaxies supported the idea that the quasars were aligned closer to the line of sight. 

\subsection{Previous tests of unified models}
\label{PS}
Barthel's proposal was not the first to suggest that different classes of AGN could be unified: for example, \cite{Orretal1982} had already explored the role of beaming in unifying flat- and steep-spectrum radio-loud quasars, and \cite{Lawrenceetal1987} had already discussed in detail possible unification of all AGN based on orientation and relativistic beaming. However, Barthel's work, and in particular the fact that it predicted different size distributions for quasars and radio galaxies, enabled new observational tests of unification. Studies such as those of \cite{Onuoraetal1991} and \cite{Gopal1992} found different linear size or angular size trends for quasars and radio galaxies. Other studies \citep[e.g.][]{Saikiaetal1994, Hongetal1995, Ubachukwuetal2002} found that the fraction of quasars present in the sample or the individual angles of the sources in the sample were consistent with Barthel's proposed scheme. Similarly, \cite{Bestetal1996, Morgantietal1997, Lahteenmetal1999}, have explored different properties of AGN and confirmed their consistency with the unified scheme. All of these studies support the idea that quasars lie at lower values of the angle to the line of sight than radio galaxies. By contrast, other studies \citep[e.g.][]{Fantietal1990, Saripallietal2005} found evidence against the unification model by studying the galaxy environments, size distribution, and or evolution of galaxies, although some of these were affected by a failure to realize that LERGs do not participate in the unified model. Most importantly in the context of this paper, a number of studies  \citep{Singaletal1993, Singaletal1996, Singaletal2013, Singaletal2014} have carried out a comparison between the expected and observed linear size ratios for the quasar and radio galaxy samples and observed discrepancies between the two. They also found varying critical angles at different redshifts (again, to some extent affected by including LERGs in their sample) and hence concluded that Barthel's unified scheme as originally proposed does not hold. A brief description of the studies and the sample sizes used by different investigations previously, either supporting Barthel's unified scheme or presenting evidence against it, are presented in Tables \ref{tablesupport} and \ref{tableagainst}.

It is important to note here that the basic assumption for the effect of projection on measured size considered in all of the studies presenting evidence against the unified models is based on a very simple model; revisiting this evidence with more physically motivated models is the goal of the present paper. In addition we note that the number of sources in samples used for testing the unification model is usually very small, and has never been more than $\sim 500$.

\begin{table*}
	\centering
	\caption{Previous studies that support Barthel's unification schemes with their respective conclusions. $N_{q}$ and $N_{rg}$ are number of quasars and number of radio galaxies (RG)} 
	\label{tablesupport}
	\begin{tabular}{m{1cm} m{1cm} m{1cm} m{4cm} m{4cm} m{4cm}}
    \toprule
		Study & $N_{q}$ & $N_{rg}$ & Test/Objective & Conclusions & Remarks \\
        \midrule
		\cite{Barthel1989} & 17 & 33 & Investigate relativistic beaming in quasi-stellar radio sources by examining large and small scale radio morphology & All radio loud quasars are beamed towards observer and powerful RGs are the unbeamed parent population & Only powerful sources are used in the sample.\\
        \addlinespace
		\cite{Onuoraetal1991} & 30 & 66 & Explore the angular diameter-redshift relation taking into account Barthel's scheme & Radio galaxies have different angular diameter-redshift relation than quasars which is an orientation effect & Non random steep spectrum bright sources ($P_{178} > 10^{26}$ W Hz$^{-1}$ sr$^{-1}$)\\
        \addlinespace
		\cite{Gopal1992} & 308 & 127 & Compare linear size and redshift of radio sources only for large sources & The evolution of linear sizes is indistinguishable for quasars and RG's & The redshift bins size is 0.1 to 2 \\
        \addlinespace
		\cite{Saikiaetal1994} & 44 & 70 & Test the unification scheme and explore the quasar fraction & Size ratios and critical angles are consistent with the unified scheme & Two redshift bins considered have linear size ratio of 0.90 and 0.45 \\
        \addlinespace
		\cite{Hongetal1995} & 41 & 2 & Investigate individual angles to line of sight and linear sizes & Individual quasars and RG's are observed at > 40$^{\circ}$ and < 50$^{\circ}$ respectively & VLBI data is used and very low number of RG's \\
        \addlinespace
		\cite{Saikiaetal1995} & 39 & 42 & Test the unification scheme and investigate the source sizes & The expected size ratio and the observed size ratio are similar & The redshift is from 0 to 2 and only bright sources are used. \\
        \addlinespace
		\cite{Ubachukwuetal2002} & 132 & 76 & Measure and compared angle to line of sight for individual sources & Radio axes of quasars are close to the line of sight and radio axes of radio galaxies are close to the plane of the sky (median of Q $\approx$ 28$^{\circ}$ and RG $\approx$ 51$^{\circ}$) & Compact steep spectrum sources are used in sample. \\
        \addlinespace
		\cite{Morabitoetal2017} & 16 & 44 & Measure critical angle and linear size ratio & The critical angle observed is 42.8$^{\circ}$ and linear size ratio is 0.32 & Sample from LOFAR for sources at power greater than $10^{25.5} WHz^{-1}$ \\
        \bottomrule
	\end{tabular}
\end{table*} 

\begin{table*}
	\centering
	\caption{Previous studies that provide evidence against Barthel's unification schemes with their respective conclusions. $N_{q}$ and $N_{rg}$ are number of quasars and number of radio galaxies (RG)} 
	\label{tableagainst}
	\begin{tabular}{m{1cm} m{1cm} m{1cm} m{4cm} m{4cm} m{4cm}}
		\hline
		Study & $N_{q}$ & $N_{rg}$ & Test/Objective & Conclusions & Remarks \\
		\hline\hline   
		\cite{Fantietal1990} & 46 & 93 & Explore properties of different classes of CSS & The difference in CSS radio galaxy and quasar morphology is not explained by projection & The linear size distributions for quasars and RG's are similar\\
		\cite{Singaletal1993} & 32 & 99 & Compare the linear size ratios for expected and observed values & Discrepancy between observed and expected size ratio for three redshift bins & Large mismatch between observed and expected size ratio at lower redshift bin\\
		\cite{Singaletal1996} & 188 & 934 & Test the unification scheme by exploring the quasar fraction for different flux densities & Systematic decrease in quasar fraction; no drop in RG and Quasar observed size ratios & Samples are from multiple surveys for decreasing flux levels\\
		\cite{Singaletal2013} & 93 & 381 & Test and observe source sizes, their flux level and redshift trends & Observed quasar sizes are not systematically smaller than RGs for same flux level and redshift & The sample (MRC) is compared with 3CRR \\
		\cite{Singaletal2014} & 45 & 85 & Test the unification scheme and investigate the source sizes and other parameters & The source sizes are not systematically smaller and there are discrepancies between observed and expected ratio & Sample is obtained for FR type-II sources \\
		\hline
	\end{tabular}
\end{table*} 

\subsection{This paper}

Our aim in the present study is to explore the unification model by studying simulated samples that assume a more physically realistic model for the projection of the lobes and take into account the known dispersion in sources' physical sizes and axial ratios. For the simulation of the effects of projection we will make use of the study conducted by \cite{Mullinetal2008}, who explored the morphological properties of Fanaroff–Riley II (FRII) quasars and radio galaxies with $z<1$. Using the study we will present a model that takes into account the effects of axial ratio in projection, which will later be used to explore test parameters such as critical angle and size ratio. We hope to answer the following questions in this study: 
\begin{enumerate}[i]
	\item What parameters need to be taken into account in modeling radio sources?
	\item What can be concluded from the simulated results of the size ratios? 
    \item What do we observe when we take resolution effects into account?
	\item What can we say about the previous studies that performed similar tests and how do they compare with our results? 
	\item What can we conclude about the validity of the unification scheme from our results and the work carried out previously by others?      
\end{enumerate}
      
In Section \ref{Sec:DataandModel} we explore the modeling in detail and outline the modeling process. In Section \ref{Sec:RnD}, we discuss the results from our study in detail, and compare our results with those of previous studies. In Section \ref{Sec:Conclusion} we summarize our results and give answers to the above questions. In this study we use a cosmology in which $H_0$ = 70 km s$^{-1}$ Mpc$^{-1}$, $\Omega_{m}$ = 0.3 and $\Omega_{\Lambda}$ = 0.7.

\section{Data and Modeling}
\label{Sec:DataandModel}

\subsection{The Cylindrical Model}
\label{Section:models}
The orientation based scheme proposed by \cite{Barthel1989} considers radio galaxies and quasars to be derived from the same AGN population, which means that they are intrinsically indistinguishable. This also means that their observed sizes are affected by geometrical projection: quasars, which are at smaller angles to the line of sight, should be systematically smaller than radio galaxies, as was indeed the case for Barthel's sample. Another aspect of the scheme is that the relative numbers of radio galaxies and quasars are dependent on the critical angle that separates quasars from radio galaxies. The unified model assumes that there is intrinsically no bias in the distribution of angles to the line of sight, so that if $\theta$ is the angle to the line of sight, then $p(\theta) = \sin\theta$. It is thus more probable for a source to be close to the plane of the sky than for it to be aligned along our line of sight. If we consider a critical angle $\theta_{c}$ such that sources are classified as quasars if $\theta < \theta_{c}$ and radiatively efficient radio galaxies otherwise, then the ratio of the numbers of quasars ($N_{Q}$) to radio galaxies ($N_{R}$) is given by:
\begin{equation} \label{Num}
	\\ \frac{N_{Q}}{N_{R}} = \frac{\int_{0}^{\theta_{c}}\sin\theta d\theta}{\int_{\theta_{c}}^{\pi/2}\sin\theta d\theta} = \frac{1-\cos\theta_{c}}{\cos\theta_{c}}
\end{equation}

This means that we can evaluate the critical angle from a given sample using the number of quasars and radio galaxies.

If radio galaxies and quasars were intrinsically all of constant length ($L_{0}$) and appeared when projected to have a length $L_{0}\sin\theta$, then the mean apparent lengths of quasars ($\bar{L}_{Q}$) would be given by: 
\begin{equation} \label{Lq}
	\\ \bar{L}_{Q} = \frac{\int_{0}^{\theta_{c}}L_{0}\sin^{2}\theta d\theta}{\int_{0}^{\theta_{c}}\sin\theta d\theta}
\end{equation}
and the mean apparent lengths of radio galaxies ($\bar{L}_{R}$) would be given by:
\begin{equation} \label{Lrg}
	\\ \bar{L}_{R} = \frac{\int_{\theta_{c}}^{\pi/2}L_{0}\sin^{2}\theta d\theta}{\int_{\theta_{c}}^{\pi/2}\sin\theta d\theta}
\end{equation}

Using Eq.(\ref{Lq}) and Eq.(\ref{Lrg}), and assuming that the distributions of intrinsic lengths are identical, we can derive the predicted size ratio between quasars and radio galaxies, obtaining: 

\begin{equation} \label{Len}
    \frac{\bar{L}_{Q}}{\bar{L}_{R}} = \frac{2 \left(\theta_C - \frac{\sin{\left(2 \theta_C \right)}}{2}\right) \cos{\left(\theta_C \right)}}{\left(\cos{\left(\theta_C \right)} - 1\right) \left(2 \theta_C - \sin{\left(2 \theta_C \right)} - \pi\right)}
\end{equation}
where the assumed constant length disappears, so that the resulting size ratio should also hold under any distribution of intrinsic lengths that is the same for radio galaxies and quasars. 

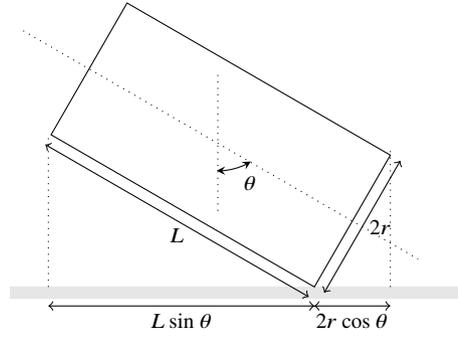
\begin{figure*}
	\begin{tikzpicture}
		% Define rectangle dimensions
		\def\width{4}
		\def\height{2}
		
		% Draw the surface plane
		\filldraw[gray!20] (0,0) -- (\width+\width/2,0) -- (\width+\width/2,-0.15) -- (0,-0.15) -- cycle;
		
		% Draw the tilted rectangle
		\begin{scope}[rotate around={-30:(\width,0)}]
			\draw (0,0) rectangle (\width,\height);
		\end{scope}
		
		% Draw the dotted line through the midpoint of the width
		\begin{scope}[rotate around={-30:(\width,0)}]
			\draw[dotted] (-\width/4, \height/2) -- (\width+\width/4, \height/2);	
		\end{scope}
		
		% projections 
		\draw[dotted] (\width/8, 0) -- (\width/8, \height);
		\draw[dotted] (\width/8+\width+\width/8, 0) -- (\width/8+\width+\width/8, \height-0.2);
		
		% Angle of tilt
		\draw[dotted] (\width/2+\width/8+0.23, 1) -- (\width/2+\width/8+0.23, \height+0.8);
		
		% Label theta between the two lines
		\draw[<->,>=stealth,rotate around={120:(2.76,1.45)}] (\width/2+\width/8+0.23, 1) arc[start angle=180, end angle=150, radius=0.9] node[midway, below right] {$\theta$};
		
		% Add arrows with labels
		\begin{scope}[rotate around={-30:(\width,0)}]
			\draw[<->] (\width, -0.15) -- (0, -0.15) node[midway, below] {$L$};
		\end{scope}
		\begin{scope}[rotate around={0:(\width,0)}]
			\draw[<->] (\width, -0.25) -- (0.5, -0.25) node[midway, below] {$L \sin\theta$};
		\end{scope}
		\begin{scope}[rotate around={150:(\width,0)}]
			\draw[<->] (\width-0.15, 0) -- (\width-0.15,-\height) node[midway, right] {$2r$};
		\end{scope}
		\begin{scope}[rotate around={180:(\width,0)}]
			\draw[<->] (\width, 0.25) -- (3, 0.25) node[midway, below] {$2r \cos\theta$};
		\end{scope}
	\end{tikzpicture}
	\caption{Representation of the cylindrical model for a source tilted at angle $\theta$ from the line of sight.}
	\label{ClynConf}
\end{figure*}

However, the above analysis, which is essentially that of \cite{Barthel1989}, assumes that radio AGN behave like one-dimensional rods (the `stick model'). Barthel argued that the lengths of radio galaxies and quasars in the 3CRR sample are in agreement with this model. In reality, however, we know that quasars and radio galaxies have a non-zero axial ratio; that is, they have finite width for a given length. This means that these galaxies would be better described as cylinders rather than rods (see Fig. \ref{ClynConf}). In this model the lobe width becomes the width of the cylinder ($2r$), where $r$ is the radius of the cylinder. The length of the source ($L_0$) becomes the height of the cylinder. It follows that the apparent length or projected length of the source is given by:

\begin{equation} \label{appLen}
	\\ L_\mathrm{proj} = L_{0} \sin\theta + 2r \cos\theta
\end{equation}

If we define axial ratio, $a = 2r/L_0$, then the mean apparent length of the quasars in the cylindrical model is given by: 

\begin{equation} \label{LqCy}
	\\ \bar{L}_{Q} = \frac{\int_{0}^{\theta_{c}} L_{0}\sin^{2}\theta d\theta + a \int_{0}^{\theta_{c}} L_{0} \cos\theta\sin\theta d\theta}{\int_{0}^{\theta_{c}}\sin\theta d\theta}
\end{equation}
whereas the mean apparent length for the radio galaxies for the cylindrical configuration is given by:

\begin{equation} \label{LrgCy}
	\\ \bar{L}_{R} = \frac{\int_{\theta_{c}}^{\pi/2} L_{0}\sin^{2}\theta d\theta + a \int_{\theta_{c}}^{\pi/2} L_{0} \cos\theta\sin\theta d\theta}{\int_{\theta_{c}}^{\pi/2}\sin\theta d\theta}
\end{equation}

\begin{figure*}
	\includegraphics[width=3in]{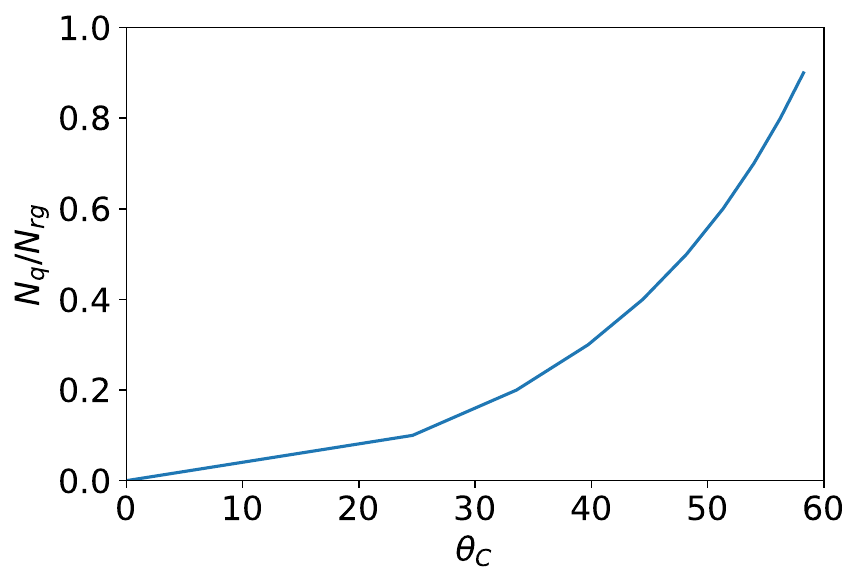}\includegraphics[width=3in]{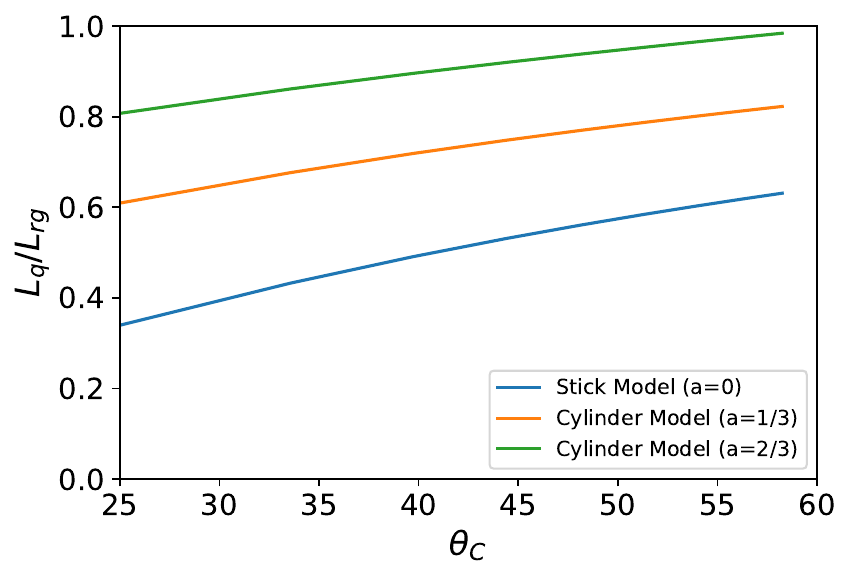}
	\caption{Quasar versus radio galaxy diagnostics in different models. The left plot shows ratio of source type numbers ratio (quasar to radio galaxy) and the right plot shows the mean length ratio (quasar to radio galaxy) as a function of the critical angle. Note the strong dependence of measured length ratio on the axial ratio.}
	\label{ratiovsangle}
\end{figure*}

Hence, the ratio of the mean of the linear sizes for the quasars and radio galaxies is given by:

\begin{equation} \label{LenCy2}
    \\ \frac{\bar{L}_{Q}}{\bar{L}_{R}} = \frac{2 \left(a \sin^{2}{\left(\theta_C \right)} + \theta_C - \frac{\sin{\left(2 \theta_C \right)}}{2}\right) \cos{\left(\theta_C \right)}}{\left(\cos{\left(\theta_C \right)} - 1\right) \left(- a \cos{\left(2 \theta_C \right)} - a + 2 \theta_C - \sin{\left(2 \theta_C \right)} - \pi\right)}
\end{equation}
Fig. \ref{ratiovsangle} shows the number ratio and length ratio for the two models for different values of the critical angle, where the axial ratio takes fixed values of 0 (stick model), 1/3, and 2/3. It can be seen that the predicted size ratios depend strongly on the value of $a$ in the population.

\subsection{Data used for modeling}
\label{sec:simulationdata}
\cite{Mullinetal2008} conducted a study on a complete sample of Fanaroff–Riley II (FRII) quasars and radio galaxies using Very Large Array (VLA) and Multi-Element Radio-Linked Interferometer (MERLIN) surveys of 3CRR sources having $z < 1$. There are 98 sources in the sample. Within the study they present the relationship between the lobe axial ratio and the largest lobe linear size of the source, observing that axial ratio depends on length, in the sense that larger sources tend to have smaller axial ratios. Using the relationship shown by \cite{Mullinetal2008}, we can simulate a sample with a similar intrinsic size and axial ratio distribution, assuming the cylindrical projection model discussed in Section \ref{ClynConf}.

\subsection{Modeling source lengths and angles}
\begin{figure}
	\includegraphics[width=3in]{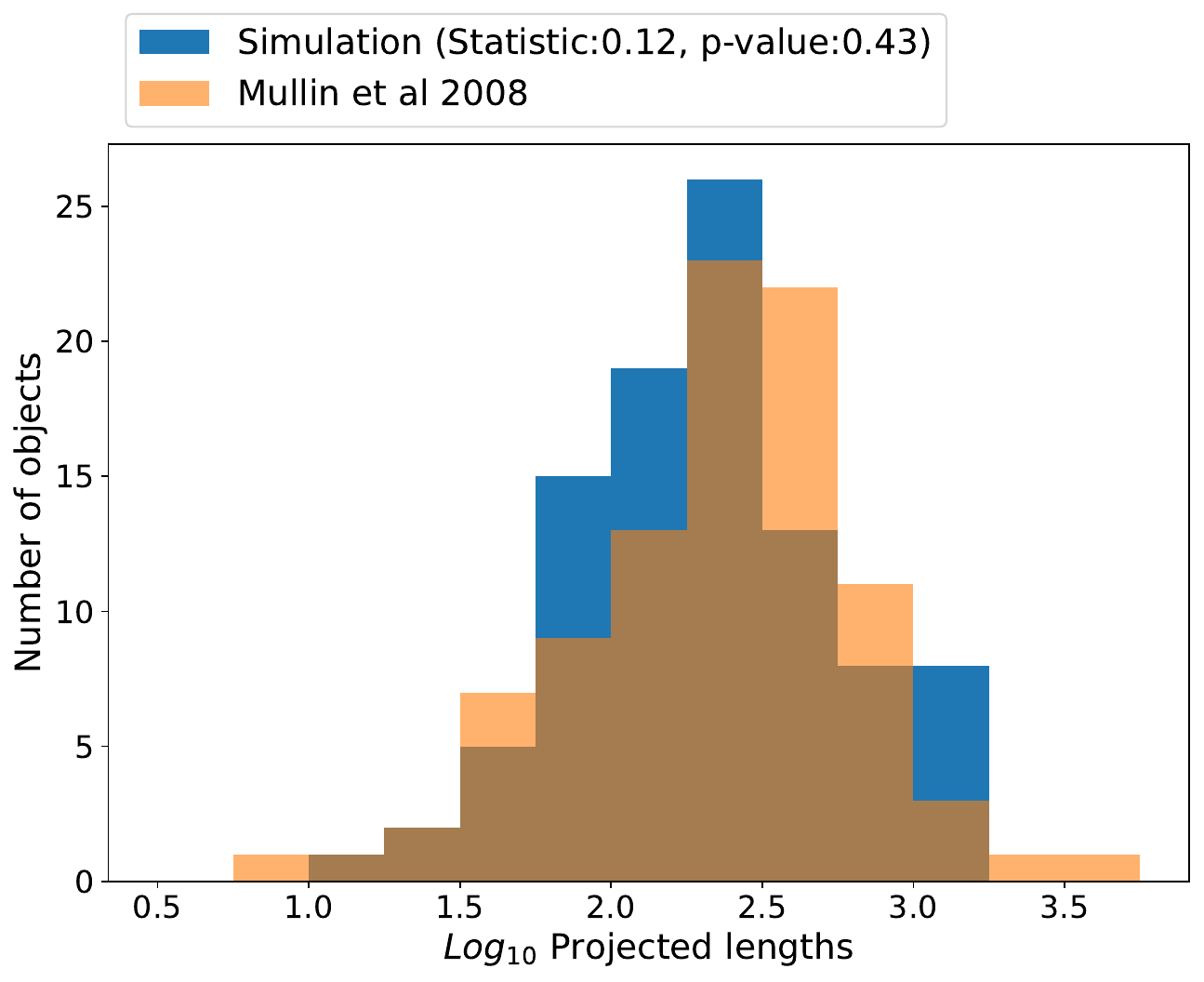}
	\caption{Distribution of projected lengths obtained for a sample of 98 sources using simulation. Overlaid is the distribution of projected lengths reported by \protect\cite{Mullinetal2008}. The legends also show the statistic and p-value obtained using the two-sample KS test.}
	\label{Fig:MullenvsLengthproj}
\end{figure}
 First we defined the intrinsic (unprojected) length of the source to be distributed log-normally, based on the fact that the maximum and minimum projected lengths of Mullin et al.'s sources lie between values close to $\sim 10$ and $\sim 2000$ kpc. We know that the cosine of the angle to the line of sight $\theta$ is uniformly distributed and so we get the distribution of $\theta$ by using the formula: $\cos\theta$ = $(1-u)$ where $u$ is a uniformly distributed variable, and this allows us to calculate projected length. We determined the parameters of the log-normal distribution ($\mu = 5.38$ log(pc), $\sigma = 1.07$) using a Kolmogorov-Smirnov (KS) test. The comparison between the distribution of projected lengths obtained using simulation and by \cite{Mullinetal2008} is given in Fig. \ref{Fig:MullenvsLengthproj}, along with KS test results. The simulated projected lengths are obtained using the true length distribution and tested using the KS test for multiple iterations. The parameters of the iteration that produce the best possible KS test statistic and $p$-value are chosen as parameters of the true length distribution. Along with the true length distribution, we also determine the axial ratio distribution simultaneously, details of which are given in the next section.

\subsection{Modeling axial ratio}
\begin{figure}
	\includegraphics[width=3in]{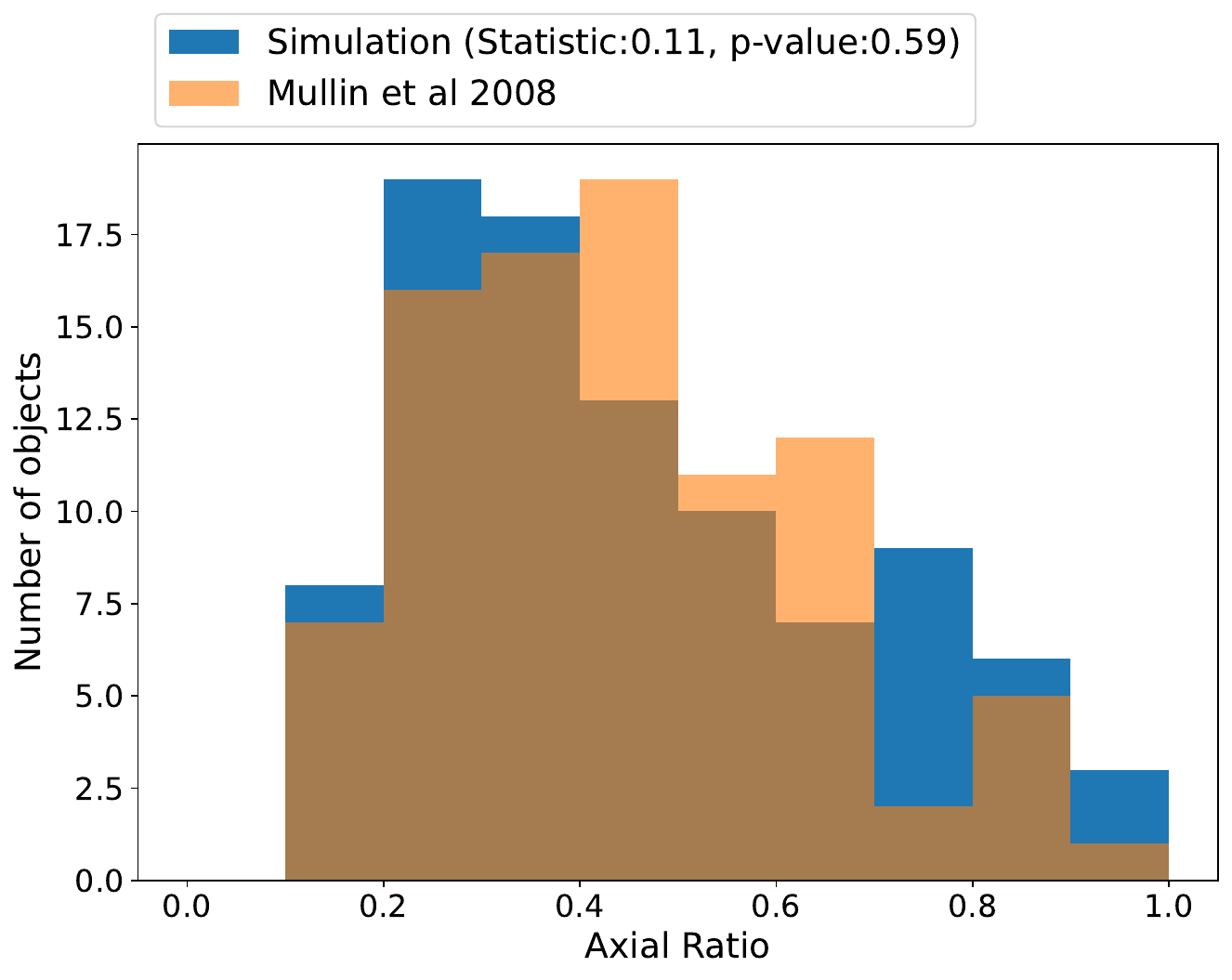}
	\caption{Distribution of projected axial ratios obtained for a sample of 98 sources using simulation. Overlaid is the distribution of projected axial ratios reported by \protect\cite{Mullinetal2008}. The legends also show the statistic and $p$-value obtained using the two-sample KS test.}
	\label{Fig:MullenvsAxialproj}
\end{figure}
We know that axial ratio cannot be a constant for a given sample and instead should follow some distribution. For this, we compare the distribution of the axial ratio versus the linear size obtained by \cite{Mullinetal2008} to the axial ratio versus the linear size plot obtained by simulation to model the distribution of the axial ratio. We can do this by adjusting the parameters of the distribution of axial ratio, such that we obtain results consistent with those from \cite{Mullinetal2008} using a Kolmogorov-Smirnov test to ensure a good match to the data. We observe a cut off in the projected axial ratio distribution less than 0.1 in the \cite{Mullinetal2008} sample, which becomes a constraint for our simulation when producing projected axial ratios. The distribution comparison from using the initial simulations and the \cite{Mullinetal2008} observed distribution shows that the actual axial ratio has a dependency on the actual source lengths, where we use the Rayleigh distribution to model its scatter. The comparison between the distribution of projected axial ratios obtained using simulation and by \cite{Mullinetal2008} is given in Fig. \ref{Fig:MullenvsAxialproj}, along with KS test results. The simulated projected axial ratios are obtained using the true axial ratio distribution and tested using the KS test for multiple iterations. Parameters of an iteration that produce the best possible KS test statistic and $p$-value are chosen as parameters of the true axial ratio distribution. The final distribution is given by; $a = (1.605 + 0.0046l)\times R(x,\sigma = 0.75)$, where $ R(x,\sigma = 0.75) $ is Rayleigh's distribution with scale factor ($\sigma$) of 0.75 and $x$ is a random value. As noted above, we truncate the intrinsic axial ratio distribution so as to avoid generating any source that has an axial ratio less than 0.1. 

\subsection{Projected lengths and axial ratio}
By using the actual axial ratio and the actual length of the source, together with the angle to the line of sight, we can simulate the projected length and projected axial ratio, where the projected length of the source is given by;
\begin{equation}
    L_{\mathrm{proj}} = aL\cos(\theta) + L\sin{\theta}
\end{equation}
where $a = \frac{2r}{L}$ is the true axial ratio, $L$ is the actual (intrinsic) length of the source, and $2r$ is the width of the lobe (cf. Eq. \ref{appLen}). 

\begin{figure}
	\includegraphics[width=\linewidth]{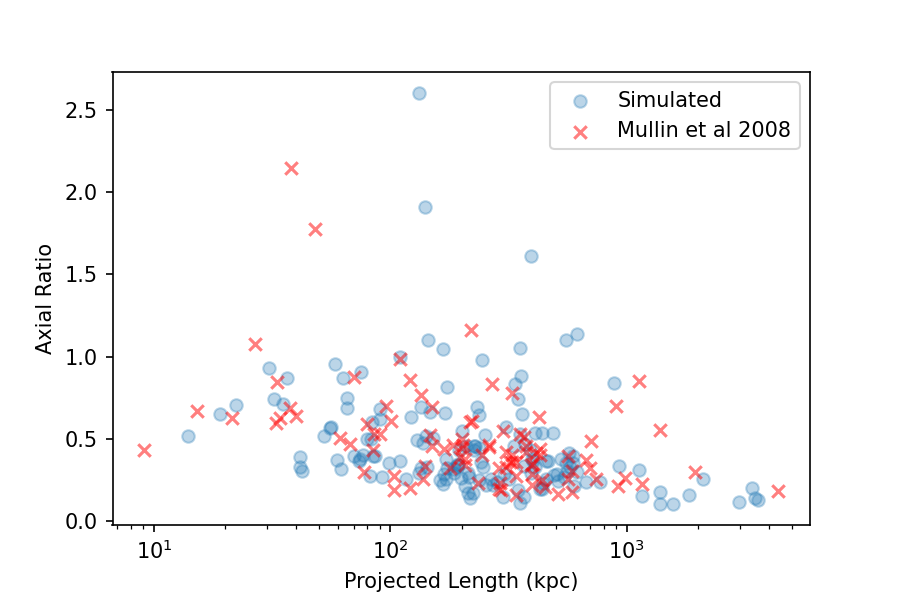}	
    \caption{Projected length of the sources versus the observed axial ratio of the source, comparing the results of our simulation with 150 sources and the distribution observed by \protect\cite{Mullinetal2008}. We see good agreement between the two distributions.}
	\label{OM}
\end{figure}

The observed distribution of lobe length and axial ratio from simulation using the above parameters for the true size and axial ratio distribution is overlaid on the \cite{Mullinetal2008} data in Fig. \ref{OM}. We use these best-fitting parameters for the intrinsic size and axial ratio distribution throughout the rest of this work. We can divide the simulated data points into quasars and radio galaxies by defining a critical angle, where the input parameters in the simulation include critical angle, the number of sources to be simulated, and number of simulation iterations. For each iteration, we generate a sample of fixed size with observed lobe lengths taking account of the projection effects in the cylindrical model and drawing the intrinsic size and axial ratio from the distributions described above. Sources are classified as radio galaxies or quasars based purely on the simulated orientation angle. The mean and median ratio of quasar and radio galaxy projected physical sizes can then be calculated.

\subsection{Resolution effects}
We also note that several surveys have a resolution limit which means that length measurements would be accurate only for a source having angular size greater than the resolution of the survey. Therefore, we also take into account in the model the resolution constraints of the survey as an additional attribute to study and compare the effects of resolution on the simulation results. For example, we can use the data from the LOFAR LoTSS deep field surveys \citep{Bestetal2023}, which provides source classifications for $\approx$ 80,000 sources. These are the deepest low-frequency surveys yet carried out, and in addition \cite{Bestetal2023} carried out spectral energy distribution (SED) fitting to the broad-band photometric data, allowing us to assess to the contribution from the AGN. The first data release (DR1) of the LoTSS Deep fields was focused on the ELAIS-N1 field, the Lockman Hole field, and the Bo\"otes field, and used only Dutch baselines, reaching an rms noise level below 20 $\mu$Jy beam$^{-1}$. The total number of sources present in the catalogue is 81,951 and a total of 14,493 objects were classified by \cite{Bestetal2023} as RLAGN. We refer to the \cite{Bestetal2023} classified LoTSS-DR1 data as LoTSS-DF from here on. We use the angular size measurements from the LoTSS-DF catalogue as a reference to get the distribution of angular sizes of sources that can act as a representative sample for recent sensitive surveys as well as a starting point to observe the effects of resolution on the unification test.

In order to take into account the resolution constraints we derived a Gaussian kernel density estimator for the redshift distribution obtained from the \cite{Mullinetal2008} catalogue, where the sampled redshift range is between 0 and 1. We can use this to resample and generate random redshift values that are drawn from the distribution seen in the \cite{Mullinetal2008} catalogue. Using the simulated redshift values we can then obtain angular sizes of sources in the simulation. Next, we reassign the angular sizes of sources smaller than 20 arcsec, below which we observe a truncated angular size distribution in the LoTSS-DF sample, and replace it with an angular size drawn from the distribution observed in the LoTSS-DF sample for unresolved sources. We repeat the process for a redshift distribution assumed to be normally distributed in the ranges between 1 and 6 with a mean of 4 and standard deviation of 1. We do this as we do not have enough data at this redshift range to get the actual redshift distribution. This range would correspond to higher redshifts missing in the \cite{Mullinetal2008} catalogue and allows us to account for the effects of higher redshifts in the analysis.

\section{Results}
\label{Sec:RnD}

\subsection{Simulation Analysis of Length Ratios}
\label{SimResults}

\begin{figure*}
	\includegraphics[width=2.1in]{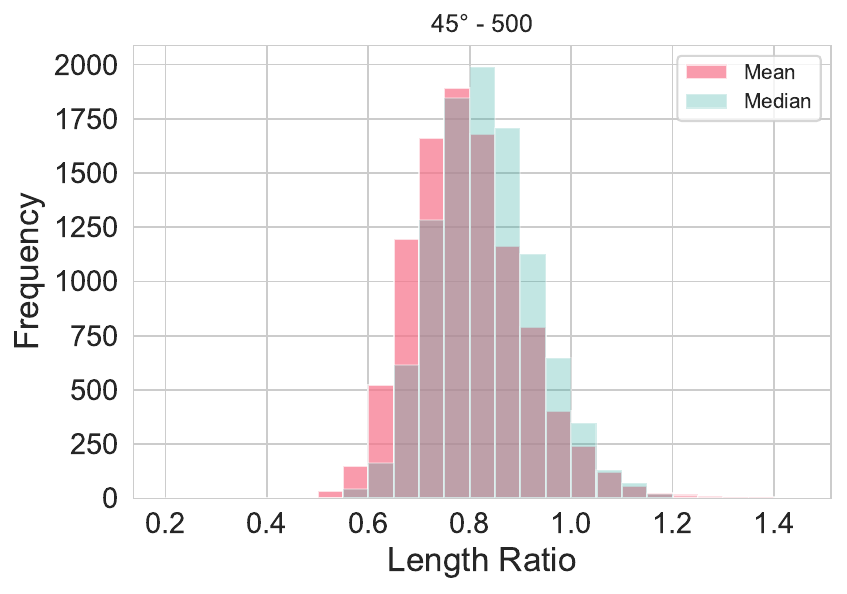}\includegraphics[width=2.1in]{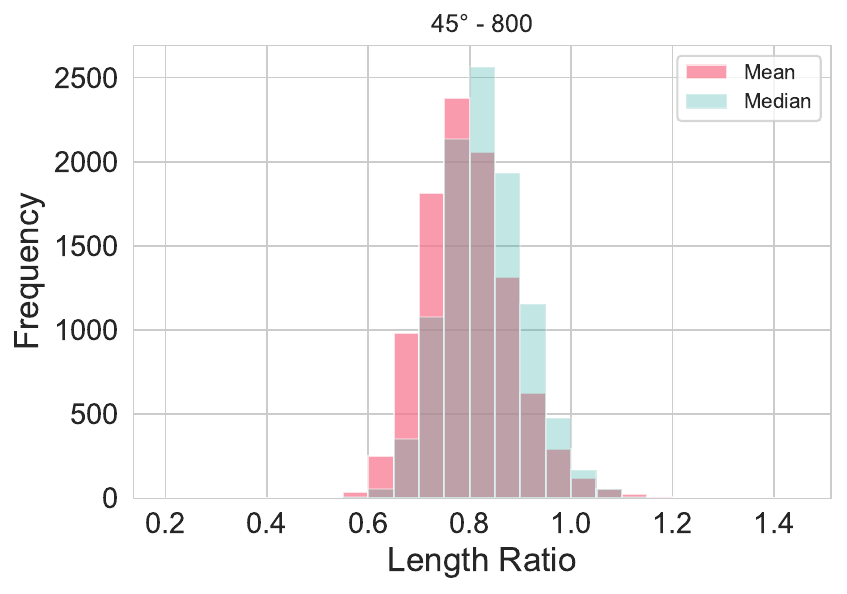}\includegraphics[width=2.1in]{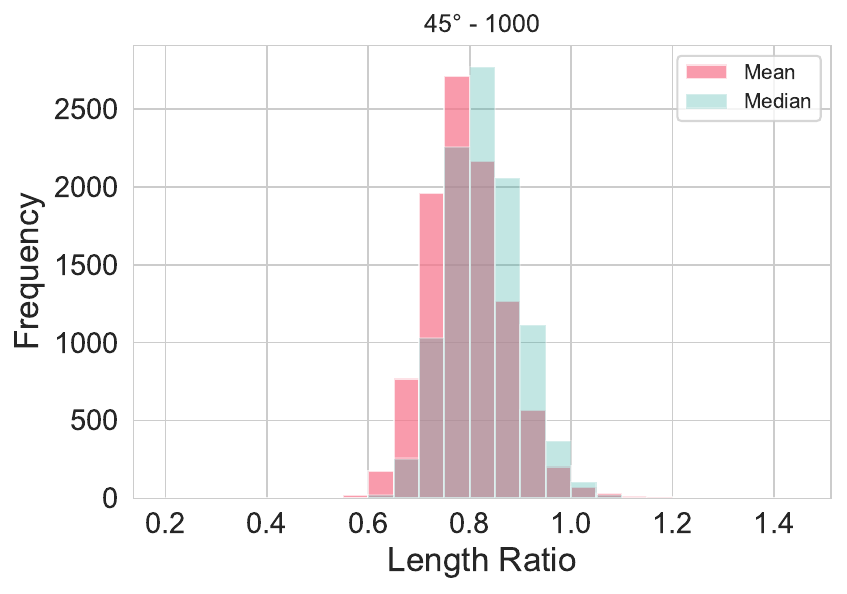}
    \includegraphics[width=2.1in]{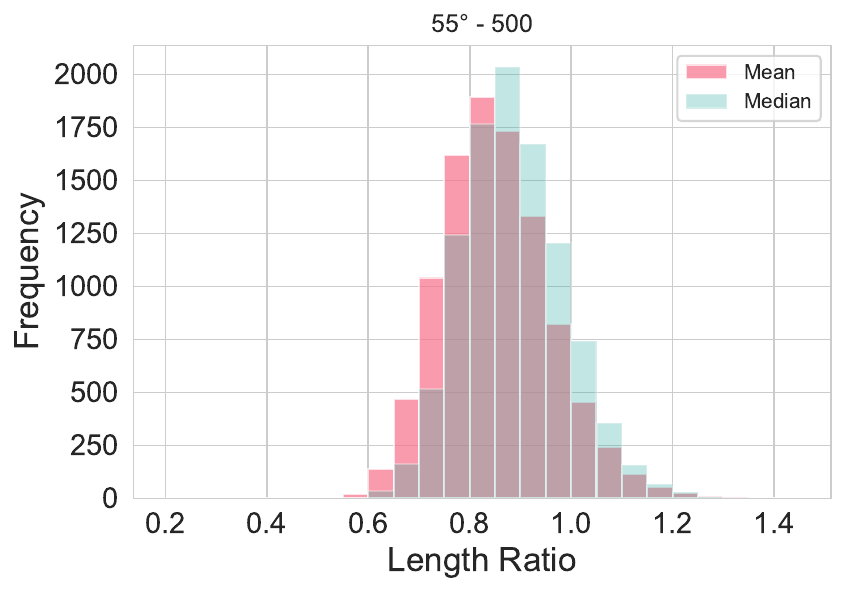}\includegraphics[width=2.1in]{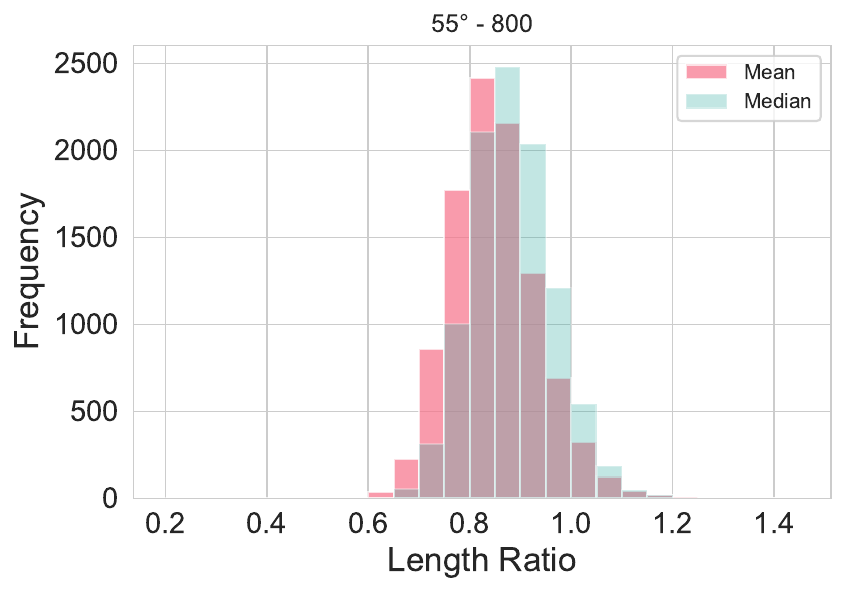}\includegraphics[width=2.1in]{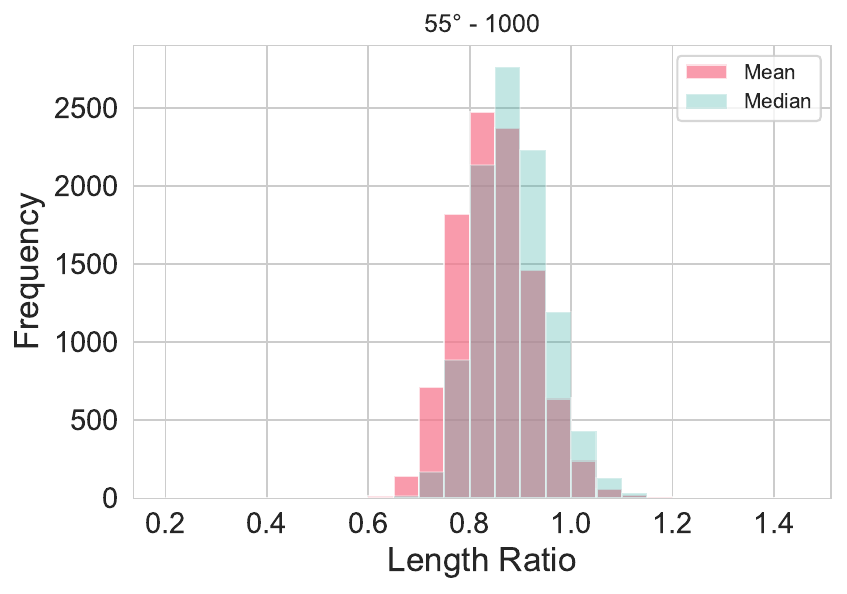}
    \includegraphics[width=2.1in]{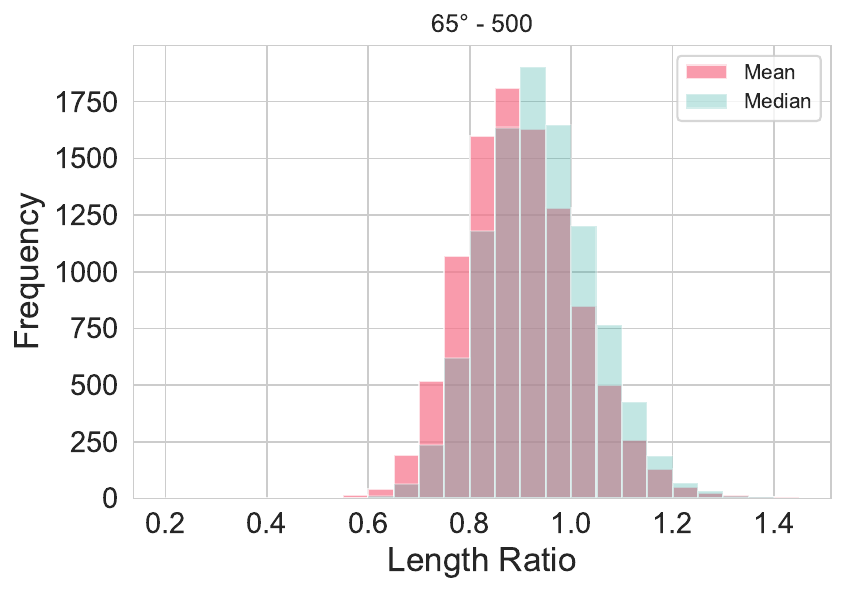}\includegraphics[width=2.1in]{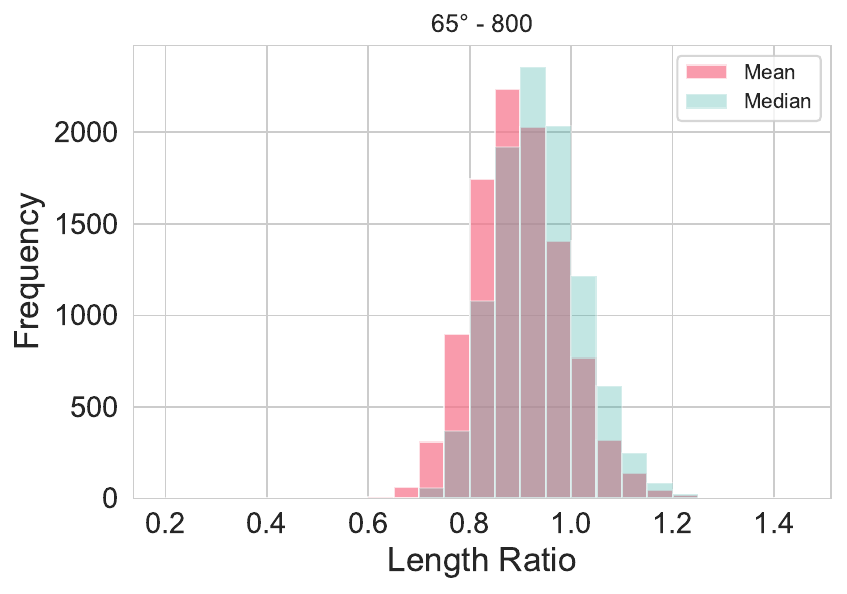}\includegraphics[width=2.1in]{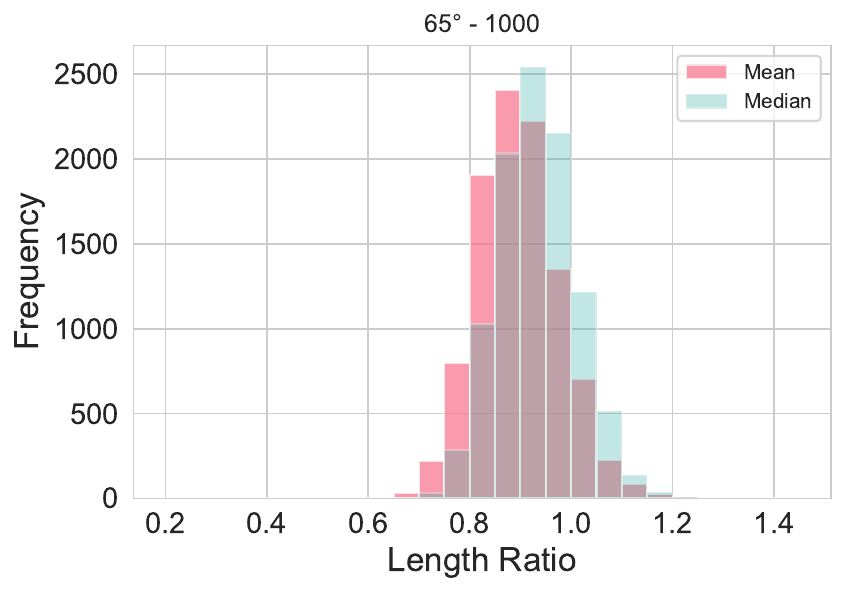}
	\caption{Distribution of ratios for average quasar linear size to average radio galaxy linear size for a sample sizes of 500, 800, and 1000 for different critical angles of 45$^{\circ}$, 55$^{\circ}$, and 65$^{\circ}$, with 10000 simulations, without resolution truncation. The three columns represent the distributions for three sample sizes (500, 800, 1000) and the three rows represent the distribution at the three critical angles (45$^{\circ}$, 55$^{\circ}$, 65$^{\circ}$). Mean values are obtained by dividing the mean lengths of quasars and radio galaxies, and the median values are obtained by dividing median lengths of quasars and radio galaxies for each iteration.}
	\label{sim2}
\end{figure*}

\begin{table*}
	\centering
	\caption{Predicted range of length ratio at 3$\sigma$ (99.86 per cent confidence) using the cylindrical model simulation for a given critical angle and 100,000 iterations. The rows give increasing sample sizes and columns show assumed critical angles, where each critical angle column reports length ratio values obtained with and without resolution truncation. For resolution truncation we draw the redshift distribution from the \protect\cite{Mullinetal2008} catalogue which has redshift range between 0 and 1.} 
    \label{Table:Sim3sigmaMullen}
	\begin{tabular}{|c|c c|c c|c c|} 
		\hline
		& \multicolumn{2}{c|}{45$^{\circ}$} & \multicolumn{2}{c|}{55$^{\circ}$} & \multicolumn{2}{c|}{65$^{\circ}$}\\
        \cline{2-7}
        \diagbox{Sample size}{Critical Angle} & Non Truncated & Truncated & Non Truncated & Truncated & Non Truncated & Truncated \\
		\hline
        300 & 0.50 - 1.36 & 0.45 - 1.41 & 0.55 - 1.41 & 0.53 - 1.45 & 0.58 - 1.51 & 0.56 - 1.55\\
        400 & 0.53 - 1.27 & 0.50 - 1.31 & 0.59 - 1.31 & 0.57 - 1.36 & 0.62 - 1.41 & 0.61 - 1.46\\
		500 & 0.56 - 1.22 & 0.53 - 1.25 & 0.61 - 1.26 & 0.60 - 1.31 & 0.64 - 1.34 & 0.63 - 1.37\\
		600 & 0.58 - 1.18 & 0.55 - 1.20 & 0.63 - 1.22 & 0.62 - 1.25 & 0.66 - 1.30 & 0.66 - 1.33\\
		700 & 0.60 - 1.15 & 0.58 - 1.17 & 0.65 - 1.19 & 0.64 - 1.23 & 0.69 - 1.27 & 0.67 - 1.31\\ 	
        800 & 0.61 - 1.12 & 0.59 - 1.15 & 0.66 - 1.17 & 0.65 - 1.20 & 0.70 - 1.24 & 0.69 - 1.28\\
        900 & 0.62 - 1.10 & 0.60 - 1.12 & 0.67 - 1.15 & 0.67 - 1.18 & 0.71 - 1.22 & 0.70 - 1.26\\
        1000 & 0.63 - 1.08 & 0.62 - 1.11 & 0.68 - 1.14 & 0.67 - 1.16 & 0.72 - 1.21 & 0.71 - 1.24\\[0.6em]
		\hline
	\end{tabular}
\end{table*}

\begin{table*}
	\centering
	\caption{Predicted range of length ratio at 3$\sigma$ (99.86 per cent confidence) using the cylindrical model simulation for a given critical angle and 100,000 iterations. The rows give increasing sample sizes and columns show assumed critical angles, where each critical angle column reports length ratio values obtained with and without resolution truncation. For resolution truncation we assume redshift distribution to be drawn from normal distribution for higher redshift ranges, between 1 and 6.} 
    \label{Table:Sim3sigmaNorm}
	\begin{tabular}{|c|c c|c c|c c|} 
		\hline
		& \multicolumn{2}{c|}{45$^{\circ}$} & \multicolumn{2}{c|}{55$^{\circ}$} & \multicolumn{2}{c|}{65$^{\circ}$}\\
        \cline{2-7}
        \diagbox{Sample size}{Critical Angle} & Non Truncated & Truncated & Non Truncated & Truncated & Non Truncated & Truncated \\
		\hline
        300 & 0.50 - 1.36 & 0.47 - 1.35 & 0.55 - 1.40 & 0.55 - 1.40 & 0.58 - 1.50 & 0.58 - 1.49\\
        400 & 0.53 - 1.27 & 0.52 - 1.28 & 0.59 - 1.31 & 0.59 - 1.33 & 0.62 - 1.41 & 0.61 - 1.40\\
		500 & 0.56 - 1.21 & 0.55 - 1.22 & 0.61 - 1.27 & 0.62 - 1.27 & 0.64 - 1.35 & 0.64 - 1.35\\
		600 & 0.58 - 1.17 & 0.57 - 1.17 & 0.63 - 1.22 & 0.63 - 1.23 & 0.66 - 1.30 & 0.67 - 1.29\\
		700 & 0.59 - 1.14 & 0.60 - 1.15 & 0.65 - 1.19 & 0.65 - 1.19 & 0.69 - 1.27 & 0.68 - 1.28\\ 	
        800 & 0.60 - 1.12 & 0.60 - 1.12 & 0.66 - 1.18 & 0.66 - 1.17 & 0.70 - 1.25 & 0.70 - 1.25\\
        900 & 0.62 - 1.11 & 0.62 - 1.10 & 0.67 - 1.15 & 0.67 - 1.16 & 0.71 - 1.22 & 0.71 - 1.22\\
        1000 & 0.63 - 1.09 & 0.63 - 1.09 & 0.68 - 1.14 & 0.68 - 1.14 & 0.72 - 1.22 & 0.71 - 1.19\\[0.6em]
		\hline
	\end{tabular}
\end{table*}

In this section we explore the distributions of length ratios observed for different sample sizes and critical angles. As discussed above, the ratio of the mean or median lengths of quasars and radio galaxies is a widely used metric for testing unified models, based on the work of \cite{Barthel1989}, since comparison between predicted values from  Eq.(\ref{Len}) with the observed lengths of the sources in any given sample can be made. The size ratio expected from the stick model measured using the equation should lie between 0.54 and 0.68 which implies that the lengths of the quasars should be around half the lengths of the radio galaxies. On the face of it, large size ratio values should be inconsistent with the simple unified model using simple length projection (the `stick model'). However, if we consider the cylindrical model for the scheme, the typical size ratios observed will always be larger than the stick model predicts, and may very well be close to unity for large axial ratios, as we showed above. In addition, we note that, for an intrinsic distribution of source sizes and axial ratios there is noise associated with any given unified model test. Even if axial ratio scatter is not accounted for and just the `stick model' is used, there would still be a range in the mean length ratio measurements  and it would be possible for a small sample to have a mean $L_Q$ greater than the mean $L_{RG}$ for example simply due to the distribution in intrinsic physical source sizes. Given that we have scatter in both physical size and axial ratio, we need to perform many iterations of Monte Carlo simulations of radio galaxy and quasar populations to account for the noise in the physical size properties of the sources. 

The unification scheme predicts that the critical angle could be somewhere near 45$^\circ$, and a reasonable starting point for simulations is a sample size of 300, since the number of sources observed are around 300-500 for the largest previous studies. Hence, we start by setting the critical angle to 45$^\circ$ and incrementing it by 10$^\circ$ for each simulation. Similarly, we start with a sample size of 300 and increment it by 100 for each critical angle, producing the median of length ratio values for 100,000 iterations and observing their distribution. For each sample size and critical angle pair we can obtain length ratio ranges at 3$\sigma$ confidence, giving the range outside which the unification model could be rejected. Our choice to run 100,000 simulations is set by the requirement to get reasonably well constrained values for the upper $3\sigma$ (99.86 per cent confidence) bound on the median length ratio on the assumption of unification. We report these values in Tables \ref{Table:Sim3sigmaMullen} and \ref{Table:Sim3sigmaNorm}, where for each critical angle there are columns showing the length ratio values. We assume that the sample is complete and contains only resolved sources, which could be useful for future survey comparisons that will produce high resolution data. In addition, in Fig.\ref{sim2}, we plot the distribution of length ratios for 10,000 iterations with the different critical angles and increasing sample sizes as described above. We also report in Tables \ref{Table:Sim3sigmaMullen} and \ref{Table:Sim3sigmaNorm}, a `Truncated' column that gives length ratios obtained after applying resolution truncation, which will be discussed in detail in Section \ref{Sec:ResolutionEffects}. In this section we only discuss the `Non-truncated' column from Tables \ref{Table:Sim3sigmaMullen} and \ref{Table:Sim3sigmaNorm}.  

We observe from the table that for the lowest critical angle of 45$^\circ$ and sample size of 300, we obtain the smallest central values of the length ratio but also the widest $3\sigma$ range of possible outcomes. As we increase the sample size for a given critical angle the range starts converging to a central value. Similarly, we see that as we increase the critical angle for a given sample size the range starts shifting to the right, i.e. we obtain much higher length ratio ranges. This is expected as we move to higher critical angles, since the number of quasars of larger sizes increases and hence so does the length ratio. From Tables \ref{Table:Sim3sigmaMullen} and \ref{Table:Sim3sigmaNorm} we can see that as the sample size increases the length ratio range converges. In addition, Fig. \ref{sim2} shows that all distributions peak at median size ratios between 0.8 and 0.9, which means that for an optimal sample size there are high chances of observing length ratios between these values. Hence, it is clear from the analysis that changing the critical angle and sample size changes the length ratio at which we can reject the unification scheme, if we use the cylindrical model.
 
The key conclusion from this section is that the ranges of length ratio that would allow a rejection of the model are different for different sample sizes and critical angles, even if the sample is unbiased and all sources are well resolved. For small numbers of sources, a very extreme quasar/radio galaxy length ratio is required to allow the unified model to be rejected at $3\sigma$ confidence. For example, for a sample size of 300 and a critical angle of $55^\circ$, a median length ratio of $>1.41$ would be required to reject the unified model, very much larger than the length ratios $\sim 1$ that earlier studies have regarded as problematic. This arises from a combination of the use of the cylindrical model and accounting for a realistic dispersion in source physical size and axial ratio.

\subsection{Effects of Resolution}
\label{Sec:ResolutionEffects}
In this section we explore the effects of observing at a finite resolution on the length ratio range. We do this to understand how the quality of the data can affect the outcome of the test. In this section we discuss the values obtained after applying the resolution truncation, using the two types of redshift distribution, assuming that the sample is complete. We report the values in Table \ref{Table:Sim3sigmaMullen} and Table \ref{Table:Sim3sigmaNorm} under the `Truncated' column for each critical angle and compare the values observed in the `Non Truncated' columns. Table \ref{Table:Sim3sigmaMullen} refers to truncated values obtained from assuming redshift distribution to follow the \cite{Mullinetal2008} sample redshift distribution, and Table \ref{Table:Sim3sigmaNorm} refers to truncated values obtained from assuming the redshift distribution to follow a normal distribution for higher redshift range. 

\subsubsection{Mullin's Redshift Distribution}
\label{Sec:ResolutionEffectsMullen}
Comparing the non-truncated and truncated values in Table \ref{Table:Sim3sigmaMullen} we can observe that the values are quite similar. There are few changes in the size ratio ranges between the two pairs of columns for each critical angle, although where they differ, it is usually in the sense that truncation makes the range of possible ratios larger. In principle, truncation should have some effect on the size ratios of these observations, but we note that the angular size distribution that we use to decide which sources should be truncated is limited by the model of source linear sizes and axial ratios, which is built by replicating data obtained from a sample of bright sources \citep{Mullinetal2008}. The 3CRR sources used by \cite{Mullinetal2008} may not be well matched to the LoTSS data, which are taken from a survey of higher sensitivity and that therefore includes many lower-luminosity sources. This is why, when we compare the fraction of objects that fall below the truncation cut in the simulation with the objects present in the LoTSS sample, that fraction is only around 20\% to 30\%. This is too low to show any major effect on the size ratios. Overall, the simulated `Truncated' observations show similar trends to the `Non Truncated' observations, with minor changes in the expected length ratio ranges. 

\subsubsection{Normal Redshift Distribution}
\label{Sec:ResolutionEffectsNorm}
Comparing the non-truncated and truncated values in Table \ref{Table:Sim3sigmaNorm} we can observe that the values are quite similar. For each critical angle we do not observe a large difference in the length ratios, similar to the observations made in Section \ref{Sec:ResolutionEffectsMullen}. The values also show convergence as we move to higher sample size for each critical angle. Overall the introduction of high redshift values in the simulation does not seem to give rise to any large deviations in the length ratios observed for source properties derived from Mullin's sample. One change we do observe, however, is that for samples derived from higher redshift the length ratios are narrower at smaller sample sizes as compared to the length ratios seen in lower redshifts. For example, at a critical angle of 45$^\circ$ for sample size of 700, Table \ref{Table:Sim3sigmaMullen} shows length ratio of 0.58 - 1.17, although a similar value is observed in Table \ref{Table:Sim3sigmaNorm} for sample size of 600 at the same critical angle. The pattern is repeated for several other values too. However, the difference in the length ratios between the samples derived at lower redshift and higher redshift is not significant enough to reach any robust conclusion about whether the use of 3CRR sources to build the model is the root cause of this difference.

\subsection{Implications for previous tests of unified models}
We have discussed previous studies of the orientation based unification scheme in Section \ref{PS}, where we briefly summarized the different conclusions presented by different investigations. We have seen that some studies present evidence in support of Barthel's unification scheme and some studies present evidence against it. However, in all of the investigations previously carried out, we can note that the number of sources used is consistently low. For almost all studies in Table \ref{tablesupport} and \ref{tableagainst}, the total number of sources is less than 200 for the entire range of redshift used. There are a few exceptions, such as the work of \cite{Gopal1992} in Table \ref{tablesupport}, \cite{Singaletal1996}, and \cite{Singaletal2013} in Table \ref{tableagainst}, where previous work has a sample size that is comparable to the sample size that we would consider appropriate to perform the unification test. However, there are other factors that need to be considered for these studies. In the study performed by \cite{Gopal1992}, the radio galaxy sample comes from a complete flux-limited catalogue, which is not true for the quasar sample, where more than half of the quasars are at low redshift ($z < 1.5$) and the sources are not homogeneously selected. For the study conducted by \cite{Singaletal1996}, the sources are homogeneously selected from different surveys, such as 3CRR, MQC, and B3, and are thus obtained by selection from different flux-limited samples. Hence, the total number of sources might be high but the comparisons are made for inhomogeneously selected sources from different surveys, which cannot be considered as a single total sample. Similarly, in the study of \cite{Singaletal2013}, the sources in the sample come from two different surveys, that is the 3CRR survey and the MRC survey. The numbers of sources {\it in each sample} from the two surveys are smaller than the number quoted in our simulation outcomes. 

Overall, previous studies have often had considerably lower sample size with respect to the sample sizes of more than 800 sources that we propose to be necessary to test for orientation based unification scheme, and/or do not use a uniform selection. From the results of the simulations described in this study, we can see that a small sample size results in a wider predicted length ratio range at $3\sigma$, and hence a higher probability of obtaining size ratios greater than 1, simply because of the random scatter imposed by the intrinsic distribution in radio galaxy linear size and axial ratio. Such results would on the face of it imply that the size of the quasars are systematically larger than radio galaxies, which would be inconsistent with a unified model, however have shown that they are not unexpected when the cylindrical projection model and a realistic axial ratio and linear size distribution are used, especially when the sample size is small. If a large homogeneously selected sample is used, then there is a low probability in the unified model of observing a large size ratio. Although as the results from simulations suggest, this might still not be enough to reject the model as the rejection ranges are different for any given critical angle and sample size. Our simulations show that the stick configuration used by many previous studies does not give reliable expected size ratio values and leads to unrealistically small expectations for the size ratio. Once we move to a cylindrical configuration and take into account a realistic distribution in source axial ratio we start to notice that the expected size ratio measurements could be much higher. This is expected for an orientation based theory because as the source, which has a cylinder-like shape, is tilted towards the observer, the diameter of the lobe will come to dominate what is seen projected on the sky. Hence we conclude that both the model of projection used and the sample size plays a pivotal role in establishing the validity of the orientation based unification scheme. As noted in the above discussion, the sample sizes used in many previous investigations are not high enough to give reliable results; for these small samples, the probability of obtaining length ratios greater than 1 simply by chance is relatively high, which means that observed length ratios greater than 1 would still not be enough to reject the unification scheme. So, in general, previous studies neither support nor rule out unified models at any high significance level. We emphasize, however, that analysis based on length ratios that purports to support the unified model, such as the original work of \cite{Barthel1989}, must also be discounted given the small sample size. So far there is no dataset which is both large and homogeneous enough to allow this particular test to be carried out.               

\section{Conclusions}
\label{Sec:Conclusion}
In this study we have used simulations to help us understand observations of the size distribution of radio galaxies and quasars. In the simulations, in contrast to earlier work, we assume that the sources have a cylindrical configuration where the length of the cylinder is the actual length of the source and the lobe width is the diameter of the cylinder. By using the \cite{Mullinetal2008} sample we construct a model that makes use of the intrinsic length and axial ratio distributions to obtain a size ratio distribution using 100,000 iterations, generating size ratios for each iteration at three different critical angles for regularly incremented sample sizes. We report $3\sigma$ confidence ranges on the size ratio values, allowing us to see what measured size ratio values would correspond to a rejection of the unified model at that confidence level. Further, we have also explored the effects of resolution on length ratio ranges and compared our results with previous investigations. Incorporating the results from the simulated data, we can answer the questions raised in Section \ref{Intro}:  
\begin{enumerate}[i]
    \item \textit{What parameters need to be taken into account in modeling radio sources?} The model requires use of intrinsic length and axial ratio distributions, where in our modeling we see that the axial ratio distribution is also dependent on the intrinsic lengths of the sources. An important caveat is that the work presented here is dependent on the sample observed by \cite{Mullinetal2008}, which is derived from a bright, flux-limited sample. Modeling using samples derived from more sensitive surveys should give us better insights into the intrinsic distributions of physical size and axial ratio for more typical objects and, hence, more accurate results.  
    \item \textit{What can be concluded from the simulated results of the size ratios?} The results from the simulation show that there is a range associated with the length ratio that can possibly be observed for any given sample at a particular critical angle, and these ratios can be greater than 1 without the possibility of rejecting the orientation based unification scheme. This means that projection effects taking into account a realistic source projection model, together with small sample sizes, could be the root cause behind observing ratios greater than 1, which has been traditionally taken as a threshold to reject the orientation based scheme. We also note that the size ratios are not constant thresholds and rather change with critical angle and sample size used. We also observe that the distributions of size ratios at different critical angles and sample sizes peak at ratios between 0.8 and 0.9, which is also higher than the size ratios of 0.5 to 0.6 that are expected to be observed for stick model. In addition, we note that as the sample size increases the size ratios converge and start becoming constant after sample size of 800. Hence, we suggest sample sizes of 800 or more sources as reasonable to conduct such analyses.    
    \item \textit{What do we observe when we take resolution effects into account?} We also explore the effects of resolution in the simulation where we reassign the angular size of small sources to be drawn from a Gaussian distribution, using LoTSS data as reference. We also include the redshift distribution from the \cite{Mullinetal2008} sample, which constrains sources at lower redshifts ($z < 1$). We also include a redshift distribution by assuming a Gaussian distribution of high redshift values in order to assess the effect of observations of a similar population of sources at higher redshift. In our modeling, inclusion of such truncation does not produce a large change in the length ratio values for both types of redshift distribution. However, we observe that the fraction of unresolved sources obtained by simulation is around 20\% to 30\%, whereas the fraction of unresolved sources in the LoTSS-DF data is around 93\%. Such a small fraction compared to the LoTSS data gives rise to the negligible change in size ratio range seen with and without resolution truncation. Hence, we conclude that for more accurate simulation results and size ratio ranges we require size and axial ratio information that is representative of LoTSS sources. In future work, we will attempt to derive this information from large numbers of LoTSS sources observed at high resolution. Our conclusions remain valid for tests of unified models based on small samples of bright sources like the 3CRR objects described by \citealt{Mullinetal2008}.
	\item \textit{What can we say about the previous studies that performed similar tests and how do they compare with our results?} Previous studies have presented statistical tests of the orientation based unification scheme, similar to the tests that we have performed, that provide evidence either against or in support of the scheme. However, an aspect common to all these studies is the use of small samples from less sensitive surveys than are now available. In addition, some use non-homogeneous selection, and some have incomplete data. These investigations have all relied on the stick model of projection to draw their conclusions, but we have argued that the cylinder model is a much more realistic representation of the sources for orientation based unification testing. Many studies have proposed that the unified model should be rejected if size ratios greater than 1 are observed, but as we have shown, this should be updated to a size ratio range that is a function of the critical angle, sample size and assumed axial ratio distribution. If we compare the results obtained from our observed data to the data used in the previous investigations, the discrepancy in the length ratios or the observation of high scattering in the critical angles is common. Our results show that when we increase the sample size and use complete samples, the chances of observing such discrepancies decreases and the size ratio range is better constrained.
    \item \textit{What can we conclude about the validity of the unification scheme from our results and the work carried out previously by others?} Overall, from the results obtained in the study we can conclude that projection effects of the intrinsic length and axial ratio distribution of the sources play a vital role in rejecting or accepting the orientation based unification scheme. The consideration that sources should be modeled as cylinders, i.e. taking into account their axial ratio distribution, makes the testing of the unification scheme more robust and reliable. This study underlines that the use of traditional parameters should be modified to include the axial ratio of sources. When this is done, there is a size ratio range associated with a certain critical angle and sample size that allows one to to accept or reject the validity of the orientation based scheme.  
\end{enumerate}
In a future paper we will investigate the constraints that can be placed on these models using large samples of radio sources derived from the LoTSS deep and wide fields.

\section*{Acknowledgments}
MJH and JCSP acknowledge support from the UK STFC [ST/Y001249/1]. FS appreciates the support of STFC [ST/Y004159/1]. 
LOFAR is the Low Frequency Array, designed and constructed by ASTRON. It has observing, data processing, and data storage facilities in several countries, which are owned by various parties (each with their own funding sources), and which are collectively operated by the ILT foundation under a joint scientific policy. The ILT resources have benefited from the following recent major funding sources: CNRS-INSU, Observatoire de Paris and Université d'Orléans, France; BMBF, MIWF-NRW, MPG, Germany; Science Foundation Ireland (SFI), Department of Business, Enterprise and Innovation (DBEI), Ireland; NWO, The Netherlands; The Science and Technology Facilities Council, UK; Ministry of Science and Higher Education, Poland; The Istituto Nazionale di Astrofisica (INAF), Italy. 

This research made use of the University of Hertfordshire
high-performance computing facility (\url{https://uhhpc.herts.ac.uk}).

%%%%%%%%%%%%%%%%%%%%%%%%%%%%%%%%%%%%%%%%%%%%%%%%%%
\section*{Data Availability}

All LOFAR data used are publicly available at \url{https://lofar-surveys.org/}.

\bibliographystyle{mnras}
\bibliography{reference2}
\end{document}